\documentclass[runningheads]{llncs}
\usepackage[T1]{fontenc}
\usepackage{newtxtext}       %
\usepackage[varvw]{newtxmath}       

\usepackage{graphicx}
\usepackage{booktabs}
\usepackage{multirow}
\begin{document}
\title{Beyond Service Discovery: Coverage Awareness in LLM-Driven Service Composition}

\titlerunning{Coverage Awareness in Service Composition}

\author{Sepideh Masoudi\inst{1}\and
Ronja Froelian\inst{1}\and
Stefan Tai\inst{1}}
%
\authorrunning{S.Masoudi et al.}

\institute{Information Systems Engineering, TU Berlin, Germany \\
\email{{smi,st}@ise.tu-berlin.de}\\
}
\maketitle              
%


\begin{abstract}
Large language models increasingly support the automated discovery and composition of services. In federated data ecosystems, this creates new opportunities for reusing existing data products and transformation services when constructing service-based data exchange modeled as data-sharing pipelines. However, finding relevant reusable assets does not imply that the available assets are sufficient to satisfy a consumer request.
In this paper, we distinguish reuse discovery, the ability to identify relevant published assets, from reuse coverage, the extent to which the available assets support the transformations required by a request. We further consider coverage awareness, the ability of a composition system to recognize insufficient coverage. We investigate this distinction using PipeFinder, an LLM-driven framework for composing data-sharing pipelines, and evaluate 99 consumer requests against a fixed catalogue of data products and transformation services. Our experiments show that relevant data products can be discovered reliably, while successful composition remains considerably more difficult. More importantly, when required transformation services are missing, the framework typically constructs a pipeline from the available services rather than recognizing that the request cannot be satisfied. Explicitly incorporating coverage checking into the planning procedure substantially improves recognition of such cases. These results suggest that coverage awareness should be treated as an explicit concern in LLM-driven service composition and cannot be assumed to follow from improved discovery or model capability.
 
\end{abstract}

\section{Introduction}
Data mesh as a service-oriented data ecosystem distributes ownership across domains, with each domain publishing \emph{data products} that encapsulate data together with the metadata and infrastructure needed to make it accessible through a service interface, and \emph{transformation services} that process and reshape the data before sharing~\cite{dehghani2022data_mesh,boris}.
When a consumer in one domain needs data held by another, service-oriented data exchange is implemented as a data-sharing pipeline.
A Data-sharing pipeline is an ordered composition of transformation services applied to one or more data products. 
In this paper, we refer to data products, their service interfaces, and transformation services collectively as \emph{published assets} in the data mesh.

One of the main values of using data mesh is its potential for reuse.
Rather than building a new data product and transformation services for each request, domain teams are expected to compose and transform what already exists.
However, realizing that expectation is difficult in practice. As the number of published assets grows,
identifying which data products are relevant, selecting appropriate transformation services, and assembling them into
an executable pipeline becomes a substantial manual effort, and reuse opportunities are missed not
because suitable assets are absent but because locating and combining them is costly and their sufficiency is hard to judge~\cite{driessen2022data,epperson2022strategies}.
Large language models (LLMs) can be an obvious choice for automating this.
Agentic systems decompose a request, select among the available assets, and chain calls to produce a result.
In existing studies, LLM-driven compositions are applied to task sets with a shared pool of available services, where the available resources are sufficient by construction, which means that there is a correct composition and the question is whether the system finds it~\cite{li2026llm,pesl2025advanced,TOSCA}. These studies primarily focus on service discovery and composition, rather than whether the available services provide sufficient coverage i.e., whether all services a request requires exist.
Moreover, they do not study the joint composition of data and services as required in data pipelines.

However, reusing existing assets to construct a new composition requires two independent conditions to be satisfied. 
The first is that the
relevant assets can be \emph{located}. 
The second is that those assets are \emph{adequate} for the
request and that the system recognizes when they are not.
In real-world data ecosystem scenarios, the second condition frequently fails to varying degrees because published assets rarely map cleanly to consumer requirements~\cite{driessen2022data,epperson2022strategies}.
Therefore, in this paper we suggest the separation of two properties of reuse-oriented composition and study them. 
\textbf{Reuse discovery} is
the ability to locate the assets relevant to a request; it is a property of retrieval.
\textbf{Reuse coverage} captures the relation between what a request requires and what the published assets provide.
A system may retrieve the right assets, but fails to assess whether they satisfy the request.

This paper makes three contributions. First, it provides an empirical study that reuse discovery and reuse coverage are decoupled concepts in LLM-driven composition (Sections~\ref{sec:res-discovery} and~\ref{sec:res-coverage}). Second, it characterizes the resulting failure as substitution rather than hallucination and shows that this failure becomes less visible as the available coverage approaches sufficiency (Section~\ref{sec:res-composition} and~\ref{sec:res-coverage}).
Third, it shows that this limitation lies in the plan generation procedure rather than being capability-bound, as it remains unchanged under metadata enrichment, self-correction, and contextual evidence, as well as with a change in the large language model used (Sections~\ref{sec:res-coverage} and~\ref{sec:res-model}).
\section{Related Work}\label{sec:related}
Repeatedly composing transformation services across data-sharing structures depends on mechanisms that support finding services and dynamically composing them on demand.
Pesl et al. automate service composition by using LLMs to create service integrations from endpoint documentation, employing RAG to pre-process API documentation and generate composition prompts that improve endpoint retrieval~\cite{pesl2025advanced}, though their focus remains on chunking and embedding strategies rather than composition itself.
Kim et al. present RESTGPT, which generates example parameter values from natural-language API descriptions to support REST API testing~\cite{kim2024leveraging}; this lays groundwork for LLM-generated API parameters but stops short of planning a chain of API calls.
Zhang et al. present a diagnostic benchmark to assess LLM hallucinations from different perspectives, including solution planning and missing-tool analysis~\cite{zhang2024toolbehonest}.
Georgievski et al. propose an agent-based system that coordinates a planner, an API grounder, and an executor to to enable service composition from natural-language requests~\cite{georgievski2026agent}.
Calvanese et al. apply the Roman model to compose process-oriented stateful services, represented as transition systems, to achieve desired client interactions~\cite{calvanese2008automatic}.

Metadata is known to enhance dataset findability, and recent work has explored generating it automatically. Singh et al. use retrieval-augmented LLMs to generate table and column descriptions for data catalogs, finding that domain teams accepted most descriptions as-is or with minor edits~\cite{singh2025leveraging}, but do not use LLMs to strengthen the semantic mapping with consumer requirements, nor measure LLM impact on this mapping.
Recent efforts within a European project suggest automating the generation of consumer-specific data products from a general one ~\cite{asg}. 
However, such generation requires a source product as input but does not support the composition of a consumer-specific data product from multiple data products and transformation steps.

Building on existing work on LLM-driven composition, this paper studies the capability of LLMs to discover relevant assets and determine whether those assets adequately satisfy a request for constructing a service-based data pipeline through reuse.

\section{Reuse Discovery and Reuse Coverage}\label{sec:concepts}

Consider a group of domains in a data mesh publishing a set of data products $\mathcal{P}$ and a set of transformation
services $\mathcal{T}$.
Having reuse-oriented data exchange means that a consumer request $q$ is satisfied by a pipeline plan including an ordered composition of services from $\mathcal{T}$ applied to one or more products from $\mathcal{P}$.
However, realizing reuse for $q$ requires two things to hold:
\emph{Reuse discovery} is the ability to locate, among the published assets, those relevant to a
request. 
Discovery is limited by how assets are described, how they are indexed,
and how the request is matched to them. 
Metadata enrichment and retrieval augmentation aim to improve the system's ability to find the right assets, which is what retrieval metrics measure.
\emph{Reuse coverage} is the relation between what a request requires and what the published
assets provide. 
When $R(q)$ denote the set of transformation services required to satisfy $q$, and
$S(\mathcal{T})$ the set of transformation services the domains provide. 
A consumer request is expressible when $R(q) \subseteq S(\mathcal{T})$; otherwise, it has a \emph{coverage gap}.
Unlike discovery, coverage is independent of how well published assets can be searched. While retrieval helps find available and relevant assets, it does not necessarily mean that those assets are sufficient.
Coverage is not binary in practice and a request can be only partially supported by reusable assets.
We quantify the degree of coverage as
the fraction of required services the domains already provide : 
\begin{equation}\label{eq:coverage}
c(q) = \frac{|R(q) \cap S(\mathcal{T})|}{|R(q)|},
\end{equation}
As a result, coverage awareness is necessary for a reuse-oriented system to recognize that a gap exists and report it, rather than composing a plan across it. This need motivates the main contribution of this paper.
\section{PipeFinder}\label{sec:framework}

\begin{figure}[t]
\centering
\includegraphics[width=1.0\textwidth]{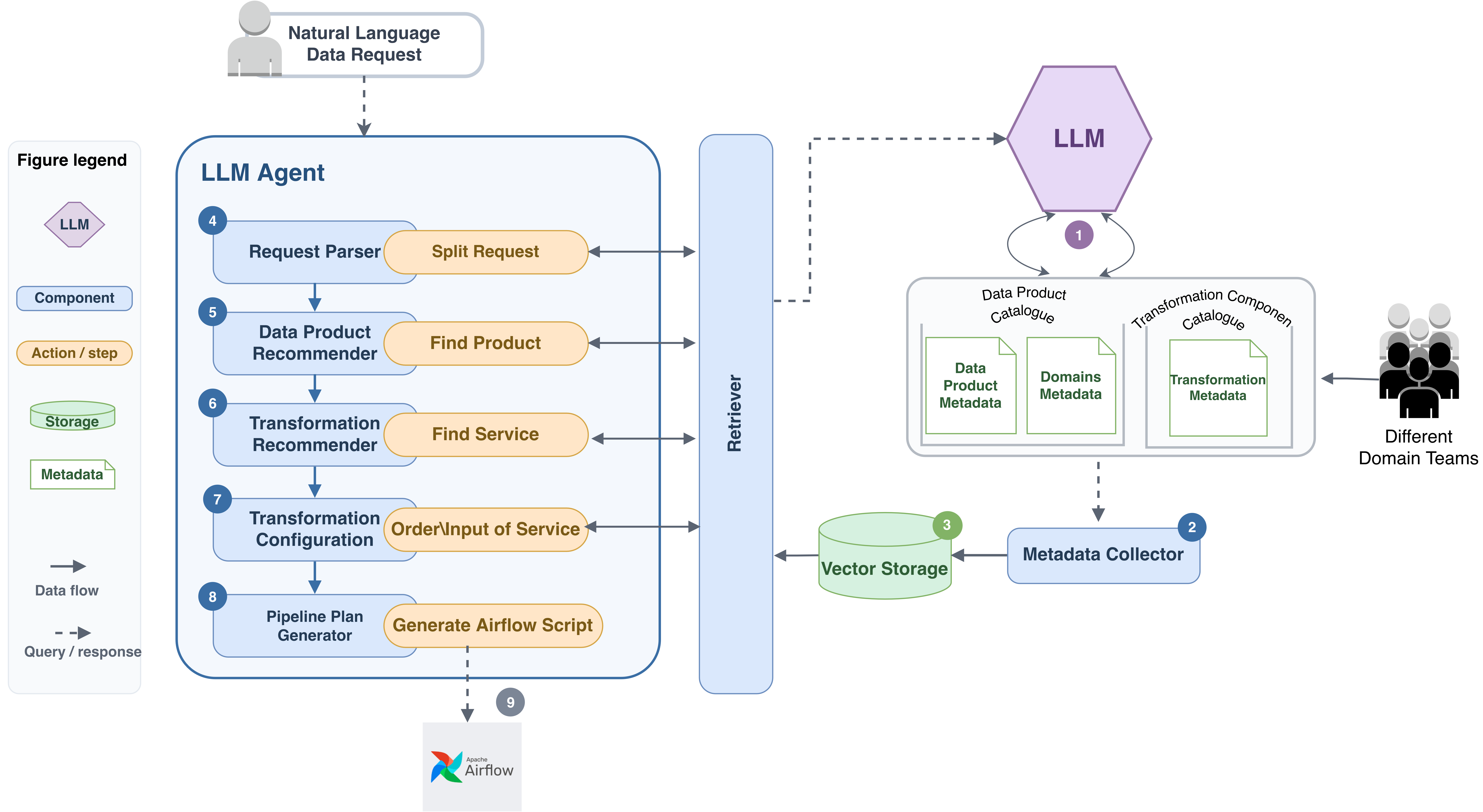}
\caption{Framework architecture of PipeFinder.}
\label{fig:arch}
\end{figure}

To study reuse discovery and coverage in LLM-driven service composition, we implement \textbf{PipeFinder}, a framework that
composes data-sharing pipelines from published data products and transformation
services~\footnote{\url{https://github.com/ISE-Research-Papers/PipeFinder}}.
It is organized into three phases, shown in
Figure~\ref{fig:arch}. The first two prepare catalogue contents for retrieval and run whenever metadata changes; the third processes requests.

PipeFinder implements the mechanisms commonly proposed for automating reuse including metadata
enrichment, the retrieval strategy, request decomposition, and iterative self-correction. Each is
implemented as a configurable option rather than a fixed design decision, so that its contribution
to reuse discovery and reuse coverage can be measured separately.

\paragraph{\textbf{Phase 1: Metadata Enrichment.}}
Shown as 1 in Figure~\ref{fig:arch}, an LLM enriches the two data product catalogue metadata supplied by domain teams, generating for each entity a
\texttt{description} that subsequently serves as retrieval context. 
\emph{Domain metadata} records
each domain's name, the products it exposes, and a description of what the domain covers;
the product list also serves as a retrieval filter. 
\emph{Data product metadata} records the
product name, the service endpoint at which it is available, and its columns; the domain team supplies the names and endpoint, and the LLM generates a description for each column and for the product as a whole. \emph{Transformation metadata} records each service's endpoint, a description of what it
computes, and the input parameters it accepts and it is generated by human.

\paragraph{\textbf{Phase 2: Retrieval.}}
Shown as 2--3 in Figure~\ref{fig:arch}, a \emph{Metadata Collector} indexes the three metadata types in a vector store, enabling
context-aware retrieval during prompt construction~\cite{lewis2020retrieval}. Entries are typed as data
product, transformation, or domain; product entries additionally record their domain, permitting
domain-scoped filtering. Three retriever subfunctions, one per type, are referred to collectively as
the \emph{Retriever}. The store is built on PostgreSQL with pgvector.

\paragraph{\textbf{Phase 3: Request Processing and Planning.}}
Shown as 4--9 in Figure~\ref{fig:arch}, a \emph{Request Parser} maps the request to available domains through identifying relevant domains and then breaking the request into steps
assigned to a domain each. 
A \emph{Data Product Recommender} identifies products within each domain
that can serve the request; \emph{Transformation Recommender} and \emph{Transformation
Configuration} select services and assign values to their parameters. 
A \emph{Pipeline Plan Generator} assembles the plan and translates it into an executable Apache Airflow DAG. Each phase is driven by calling LLM with a high-level prompt completed with metadata fetched from the vector store, so users do not need to author prompts per domain or request.
The framework is independent of a data product's internal format provided it is exposed through a
RESTful endpoint returning JSON.
\section{Evaluation Methodology}\label{sec:method}

\subsection{Experimental Setup}
The federation is constructed from the BIRD Mini-Dev benchmark~\cite{bird}. 
Each of its 11
databases is treated as a domain, since each covers a coherent subject area, and the 75 tables
across them are exposed as data products, each wrapped in a service serving its data over a REST
endpoint.

Transformation services are implemented as parameterized RESTful services. The catalogue comprises
eleven: \texttt{retrieve} and \texttt{combine} for obtaining and joining products; \texttt{filter}
and \texttt{sortby} for selection and ordering; \texttt{sum}, \texttt{mean}, \texttt{max},
\texttt{min} and \texttt{count} for aggregation; \texttt{divide} for ratios; and
\texttt{returnResult} for delivering output.
Grouping, deduplication and limiting are excluded as service parameters rather than services.

The natural-language formulation provided by the BIRD Mini-Dev query serves as the consumer request, and its reference SQL is parsed to get the transformations the request requires. 
We parse each request's reference SQL with \texttt{sqlglot}, producing an abstract syntax tree, and search it for the relational operations the query performs.
Each detected operation is mapped to the transformation catalogue, yielding the required operations, the operations covered by available services, and the remaining operations.
This gives three attributes per request about whether it is expressible or has a coverage gap; the
partial-coverage fraction $c(q)$ of Equation~\ref{eq:coverage}; and which operations are missing when some are. 
Across the benchmark five operations fall outside the catalogue including addition,
subtraction, multiplication, conditional expressions, and subqueries. 
The same parse applies the
ground truth for discovery. 
The data products a request requires are the tables its query reads. 

From the 499 benchmark queries we draw a
stratified sample of 99 as the evaluated requests.
The evaluated sample comprises 63 expressible requests and 36 with a coverage gap, spanning the benchmark's three difficulty levels (29 simple, 50 moderate, 20 challenging). 
A correct plan must draw on two or more data products for 80\,\% of them, placing the evaluation in the multi data product rather than in single data product retrieval.
In Phase 1, GPT-4o was used for metadata enrichment to ensure model diversity, as the retriever uses two versions of Gemini, as explained below.
All vector stores use \texttt{text-embedding-004}. By default, the planner does not offer a way to report unfulfillable requests.
\subsection{Measuring Reuse Discovery}
Discovery is measured over the retrieved set. For each request the five highest-ranked retrieved
assets are compared against the reference assets, and we report hit rate (at least one required
asset retrieved), mean reciprocal rank (the reciprocal of the position of the first), and exact
match (all required assets retrieved), defined in Equation~\ref{eq:metrics}.

\begin{equation}\label{eq:metrics}
\begin{aligned}
\text{HR} &= \tfrac{1}{|Q|}\textstyle\sum_i \mathbb{1}[\mathcal{R}_i \cap \mathcal{G}_i \neq \emptyset], \quad
\text{MRR} = \tfrac{1}{|Q|} \textstyle\sum_i \tfrac{1}{\text{rank}_i}, \\
\text{EM} &= \tfrac{1}{|Q|}\textstyle\sum_i \mathbb{1}[\mathcal{G}_i \subseteq \mathcal{R}_i]
\end{aligned}
\end{equation}

Here $\mathcal{R}_i$ is the set retrieved for request $i$, $\mathcal{G}_i$ the set a correct plan
requires, and $\text{rank}_i$ the position of the first element of $\mathcal{G}_i$ in
$\mathcal{R}_i$; a request retrieving no required asset contributes zero to MRR. 

\subsection{Measuring Reuse Coverage}
Coverage is measured on the same 99 requests, in three complementary ways.
First, generated plans are executed and their output compared row by row against the reference result. 
A plan is correct only when the output matches exactly; we call this \emph{execution accuracy}.  
Second, for coverage-gap requests the outcome is further classified by how the attempt failed. 
A plan that fails to execute signals its own failure; one that executes and returns incorrect data does not. 
Classifying failures this way reveals whether the visibility of failure depends on how
nearly the request was covered.

Third, we add a diagnostic in which the planner is given an inventory of available
operations and instructed to report infeasibility rather than compose a plan when the request cannot be satisfied. It is evaluated on all 36 coverage-gap requests, on which the framework should refrain, and on 20 of the 63 expressible requests, on which it should not. 
The 20 comprise 14 moderate and 6 challenging requests and include none of the 26 simple ones, so they represent the harder end of the expressible set. 
We additionally record which missing operation the framework names when it
refrains, distinguishing recognizing that a gap exists from localizing where it lies.

\subsection{Experimental Conditions}\label{sec:conditions}
For discovery, we cross the \emph{retrieval strategy}: dense retrieval over the vector store;
a hybrid ensemble combining dense retrieval with a BM25 sparse retriever; or a hierarchical strategy that identifies the relevant domain from a
collection-level index before retrieving within it.
The \emph{metadata enrichment} is performed as two parallel vector stores. The enriched store contains the
generated descriptions; the unenriched store contains only the metadata of Section~\ref{sec:framework}.  
Both are built by the same indexing code from the same catalogue, and therefore with
the same embedding model, so the presence of generated descriptions is the only difference.

For composition and coverage, we consider all combinations of three factors:
\emph{metadata enrichment}; \emph{iterative self-correction}, in which a generated plan is returned
to the model for refinement over up to three additional rounds; and \emph{evidence}, in which the
request is accompanied by the benchmark's explanatory annotation clarifying how request terms relate
to schema elements.
Gemini 3.6 Flash is used as the default model. To separate limitations of the mechanisms from
limitations of model capability, the principal conditions are replicated on Gemini 3.1 Pro, a more capable model of the same family.

\section{Evaluation Results}\label{sec:results}

\subsection{Reuse Discovery}\label{sec:res-discovery}

\begin{figure}[t]
\centering
\includegraphics[width=0.5\textwidth]{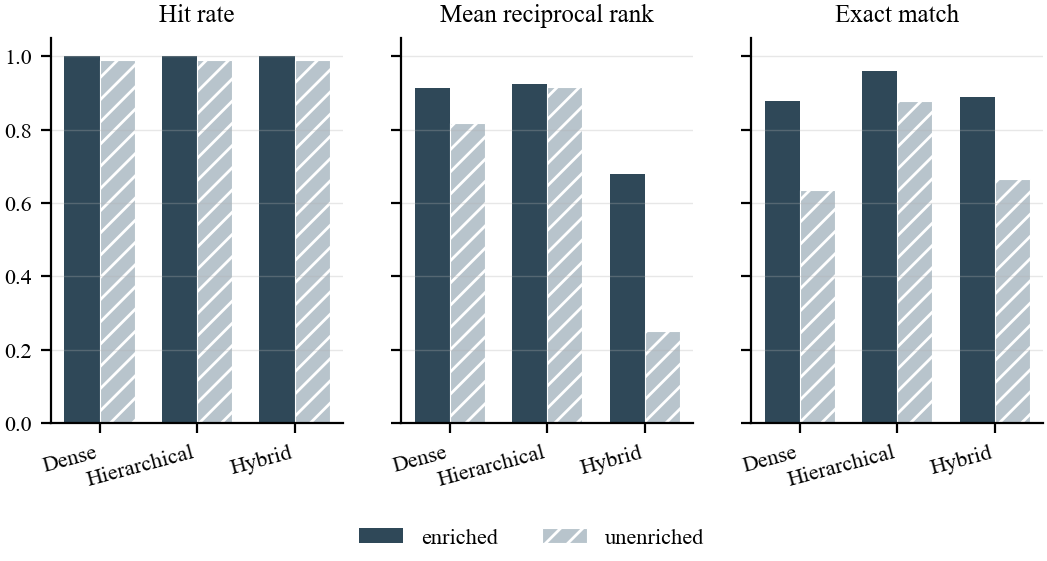}
\caption{Data-product discovery achieves consistently high hit rates, while metadata enrichment mainly affects ranking under lexical hybrid retrieval ($n=99$).}
\label{fig:disc}
\end{figure}

\begin{table}[t]
\centering
\caption{Reuse discovery for data products ($n=99$).}
\label{tab:disc}
\begin{tabular}{llccc}
\toprule
Retrieval & Catalogue & HR & MRR & EM \\
\midrule
Dense        & enriched   & 1.000 & 0.914 & 0.879 \\
Dense        & unenriched & 0.990 & 0.819 & 0.636 \\
\addlinespace
Hierarchical & enriched   & 1.000 & 0.924 & 0.960 \\
Hierarchical & unenriched & 0.990 & 0.917 & 0.879 \\
\addlinespace
Hybrid       & enriched   & 1.000 & 0.680 & 0.889 \\
Hybrid       & unenriched & 0.990 & 0.250 & 0.667 \\
\bottomrule
\end{tabular}
\end{table}

Table~\ref{tab:disc} and Figure~\ref{fig:disc} show three main observations. First, the data products required by a correct plan are reachable in every case.
Hit rate is 1.000 under all three strategies with the enriched catalogue and 0.990
without it. 
Second, the retrieval strategy determines how highly they are ranked.
Hierarchical and dense retrieval are close (MRR 0.924 and 0.914) and both well ahead of the hybrid ensemble (0.680); hierarchical attains the highest exact match, 0.960, meaning that for 95 of 99
requests every required product appeared among the five retrieved. 
Third, the benefit of
enrichment depends on the retrieval strategy.
Generated descriptions are worth $+0.007$ MRR under hierarchical retrieval and $+0.095$ under dense retrieval, but $+0.430$ under the hybrid ensemble.

Discovery of transformation services (Table~\ref{tab:disc-serv}) shows a different balance.
Enrichment now improves ranking under both strategies, by $+0.103$ and $+0.123$.
Since the service metadata is complicated, a description adds more information. 
Exact match, however, is low
throughout. 
Requests require between 2 and 7 services, and 11 of the 99 require more than
the 5 retrieval returns; restricting to the 88 requiring 5 or fewer raises exact match only to 0.091 and 0.182. 
Therefore, failure to retrieve the complete set of services a plan requires is a real limitation rather than a consequence of retrieval depth.

\begin{table}[t]
\centering
\caption{Reuse discovery for transformation services ($n=99$).}
\label{tab:disc-serv}
\begin{tabular}{llccc}
\toprule
Retrieval & Catalogue & HR & MRR & EM \\
\midrule
Dense  & enriched   & 0.960 & 0.747 & 0.081 \\
Dense  & unenriched & 0.960 & 0.644 & 0.051 \\
\addlinespace
Hybrid & enriched   & 1.000 & 0.613 & 0.162 \\
Hybrid & unenriched & 0.990 & 0.490 & 0.071 \\
\bottomrule
\end{tabular}
\end{table}

\subsection{Composition of Discovered Assets}\label{sec:res-composition}

\begin{figure}[t]
\centering
\includegraphics[width=0.5\textwidth]{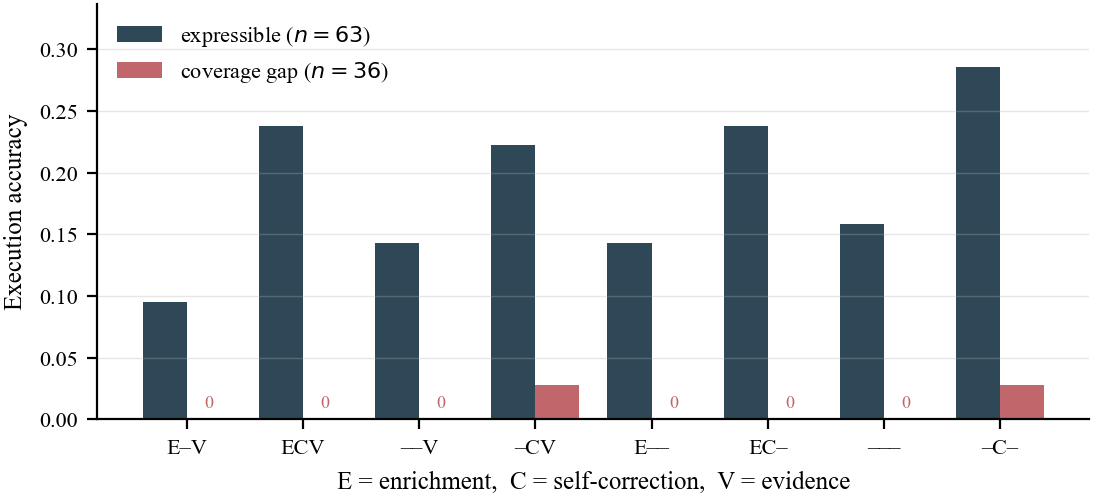}
\caption{Execution accuracy across the eight composition conditions.}
\label{fig:comp}
\end{figure}

Table~\ref{tab:comp} and Figure~\ref{fig:comp} report execution accuracy across the eight
conditions crossing enrichment, self-correction and evidence, separated by coverage class.

\begin{table}[t]
\centering
\caption{Execution accuracy ($n=99$). E~=~enrichment, C~=~self-correction,
V~=~evidence.}
\label{tab:comp}
\begin{tabular}{cccccc}
\toprule
E & C & V & Overall & Expressible & Gap \\
\midrule
\checkmark & --         & \checkmark & 0.061 & 0.095 & 0.000 \\
\checkmark & \checkmark & \checkmark & 0.152 & 0.238 & 0.000 \\
--         & --         & \checkmark & 0.091 & 0.143 & 0.000 \\
--         & \checkmark & \checkmark & 0.152 & 0.222 & 0.028 \\
\checkmark & --         & --         & 0.091 & 0.143 & 0.000 \\
\checkmark & \checkmark & --         & 0.152 & 0.238 & 0.000 \\
--         & --         & --         & 0.101 & 0.159 & 0.000 \\
--         & \checkmark & --         & 0.192 & 0.286 & 0.028 \\
\bottomrule
\end{tabular}
\end{table}

On expressible requests, the three mechanisms behave differently. 
\emph{Self-correction} raises accuracy in all three comparisons in
which it is the only factor varied, by $+0.079$, $+0.095$ and $+0.127$; at $n=63$ these correspond
to five, six and eight additional requests, and the Wilson intervals of the compared conditions
overlap in every case (for the largest, $0.159$ [0.089, 0.268] against $0.286$ [0.189, 0.407]). 
\emph{Enrichment} produces
changes of $-0.016$, $+0.016$ and $+0.048$, and \emph{evidence} changes of $0.000$, $+0.016$ and
$+0.063$; both straddle zero and neither shows a consistent sign. 
Neither the descriptions that improve
discovery nor the annotations clarifying how request terms map to schema elements measurably
improve the framework's ability to assemble a correct plan from assets it has already found.

The mechanisms also differ in cost. Averaged over the 99 requests, planning takes 97\,s per request
with no mechanism enabled, and 99\,s and 105\,s with enrichment or evidence alone. Self-correction
raises this to between 129\,s and 135\,s, an increase of 25--38\,\%, and the four conditions
applying it are the four slowest of the eight. The only mechanism that improves composition is
therefore the only one carrying a substantial cost.

Absolute accuracy on expressible requests remains low, between 0.095 and 0.286. Accuracy in the
best condition is 0.286 on requests drawing on one product and 0.308 on two, falling to 0.111 on
three; across the benchmark's own difficulty labels the variation is smaller (0.346 simple, 0.226
moderate, 0.333 challenging). 
The number of products a plan must join predicts failure better than
the difficulty of the underlying query.

\subsection{Reuse Coverage}\label{sec:res-coverage}

In our setup, a coverage gap always arises from a missing transformation service, never from a missing data product, because all user requests in the selected benchmark can be answered using the provided data by the benchmark.
As shown in Table~\ref{tab:comp}, execution accuracy on coverage-gap requests remains near zero across all eight conditions. It is 0.000 in six conditions and 0.028 in the remaining two, corresponding to only one correct request out of 36. 
Thus, none of the tested configurations produces a different outcome for requests with a coverage gap.
Therefore, in none of the eight conditions was any request reported as unsatisfiable. For every one of the 36 requests the catalogue cannot serve, in every condition, the framework emitted a pipeline plan. 
Pooled across conditions this is 0 reports of infeasibility in 288 request-condition
pairs, bounding the true rate at 1.0\,\% or below with 95\,\% confidence, and 2 correct outcomes in 288 (0.7\,\%, 95\,\% CI [0.2\,\%, 2.5\,\%]). 

An important observation is that the plans produced for these requests do not invoke services that do not exist. 
Across the 36 coverage-gap requests in the enriched, self-corrected, evidence-supplied condition, every service invoked is one the catalogue provides and no invented endpoint appears in any plan.
What the plans do instead is fall short. A coverage-gap request requires 5.58 operations on average, and the plans generated for them invoke 3.67 services which is fewer than the 4.37 invoked for expressible requests.  
The framework assembles a shorter plan from existing services and returns it as though it answered the request. For example, a subtraction request may be served by a plan that retrieves, filters, and sums, quietly answering a different question.

\begin{figure}[t]
\centering
\includegraphics[width=0.8\textwidth]{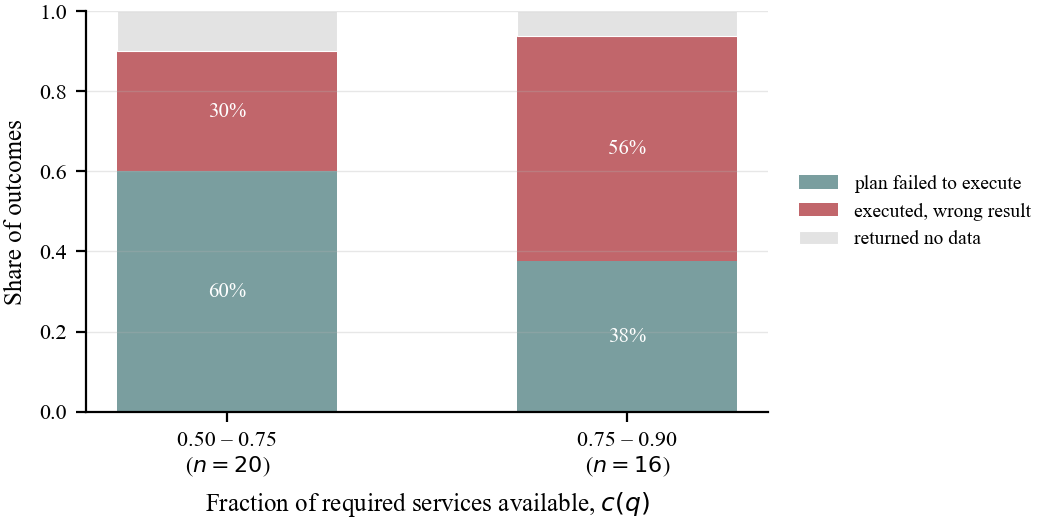}
\caption{Failure mode on coverage-gap requests, stratified by partial coverage.}
\label{fig:danger}
\end{figure}

Moreover, coverage-gap requests are not uniformly infeasible. The mean partial-coverage fraction is 0.759, and
72\,\% of them have at least three quarters of their required transformation services available. 
Stratifying by
$c(q)$ shows the mode of failure changes even though the rate does not
(Figure~\ref{fig:danger}). 
Where the catalogue supplies between half and three quarters of the
required transformation services, 60\,\% of attempts break on execution and 30\,\% execute and return incorrect data. Where it supplies more than three quarters, 37\,\% break and 56\,\% execute and return incorrect data.
Therefore, the closer the published services come to sufficiency, the more smoothly the framework bridges the remaining gap, and the more likely the resulting pipeline is to fail silently.

In the diagnostic condition the planner is given the inventory of available operations and
instructed to report infeasibility rather than compose a plan (Table~\ref{tab:elicit}).

\begin{table}[t]
\centering
\caption{Outcomes when infeasibility is offered as an explicit alternative ($n=56$).}
\label{tab:elicit}
\begin{tabular}{lcc}
\toprule
Request class & Reported infeasible & Produced a plan \\
\midrule
Coverage gap ($n{=}36$) & \textbf{33} & 3 \\
Expressible ($n{=}20$)  & 4 & \textbf{16} \\
\bottomrule
\end{tabular}
\end{table}

The framework refrains on 33 of the 36 requests it cannot satisfy (91.7\,\%, 95\,\% CI [78\,\%,
97\,\%]) while serving 16 of the 20 expressible requests in the control set (80.0\,\%). 
The control is small and its interval correspondingly wide (unnecessary refusal 20.0\,\%, 95\,\% CI
[8\,\%, 42\,\%]), so it establishes that the framework does not refuse indiscriminately but does
not pin down the rate at which it refuses unnecessarily. Recognizing the coverage boundary is
therefore within reach; it is simply not part of what the default procedure asks for.
The distance between 0.000 correct outcomes by default and 91.7\,\% correct refusal when asked is a
property of how the planning procedure is structured, not of the model's capability.

However, two points should be noted.
First, one in five expressible requests is refused, so eliciting coverage awareness trades unnecessary refusals against undetected failures.
Second, recognizing that a request is infeasible is more reliable than identifying which transformation is missing. 
The named transformation corresponds to the one actually missing in 75.8\,\% of cases. Refusal rates remain similar across partial-coverage strata (90.0\,\% and 93.8\,\%). Thus, the near-complete cases with the highest rate of silent failures are not harder to recognize when the missing transformation is explicitly queried.
Adding self-correction, refusal moves from 91.7\,\% to 86.1\,\%
and unnecessary refusal from 20.0\,\% to 15.0\,\%, both within sampling range. 

\subsection{Model Capability}\label{sec:res-model}
\begin{table}[t]
\centering
\caption{Execution accuracy difference using Gemini 3.1 Pro vs. Gemini 3.6 Flash.}
\label{tab:models}
\begin{tabular}{lcccc}
\toprule
Condition & Model & Overall & Expressible & Gap \\
\midrule
\multirow{2}{*}{E\,--\,V} & smaller & 0.061 & 0.095 & 0.000 \\
                          & larger  & 0.141 & 0.222 & 0.000 \\
\addlinespace
\multirow{2}{*}{E\,C\,V}  & smaller & 0.152 & 0.238 & 0.000 \\
                          & larger  & 0.152 & 0.238 & 0.000 \\
\addlinespace
\multirow{2}{*}{--\,--\,V}& smaller & 0.091 & 0.143 & 0.000 \\
                          & larger  & 0.121 & 0.190 & 0.000 \\
\addlinespace
\multirow{2}{*}{E\,--\,--}& smaller & 0.091 & 0.143 & 0.000 \\
                          & larger  & 0.131 & 0.206 & 0.000 \\
\bottomrule
\end{tabular}
\end{table}
The principal conditions were also replicated with the more capable Gemini 3.1 Pro.
As shown in Table~\ref{tab:models}, the larger model composes expressible requests more accurately in three of four conditions, most
markedly in the uncorrected baseline where accuracy rises from 0.095 to 0.222. Where self-correction
is already applied the two converge exactly at 0.238, therefore, model capability and iterative correction
substitute for one another. 
On the other two properties the larger model changes nothing. Discovery
is marginally worse (MRR under hierarchical retrieval falls from 0.882 to 0.837), consistent with
discovery being determined by retrieval rather than by the model consuming the retrieved context.
Execution accuracy on coverage-gap requests remains exactly 0.000 in every replicated condition, and
no such request was reported as unsatisfiable.
Model capability therefore improves only assembling a plan for a request that is both discoverable and covered but does not improve coverage awareness.
\section{Conclusion}\label{sec:conclusion}

This paper separated reuse discovery from reuse coverage in LLM-driven composition of data-sharing pipelines and studied both empirically. 
Our results indicate that reuse discovery and reuse coverage must be evaluated and engineered separately. Retrieval quality is not a proxy for whether a request can be satisfied, and composition capability does not entail knowing the limits of what has been composed. For reuse-oriented data-sharing pipelines, coverage awareness needs to be built into the planning procedure as an explicit check rather than assumed to emerge from discovery or model capability. Future work includes examining coverage awareness under partial or evolving catalogues, and studying mechanisms that reduce the trade-off between missed gaps and unnecessary refusals.
\begin{credits}
\subsubsection{\ackname} 
Claude Opus was used for text grammatical and phrasing improvements. This work was funded by BMFTR under the GANGES2 project (16KIS2609). The views expressed are those of the authors and do not necessarily reflect those of the funding authority.
\end{credits}
%
%
%
\bibliographystyle{splncs04}
\bibliography{ref-short}
\end{document}